

\input lanlmac
\input amssym
\input epsf

\newcount\figno
\figno=0
\def\fig#1#2#3{
\par\begingroup\parindent=0pt\leftskip=1cm\rightskip=1cm\parindent=0pt
\baselineskip=13pt
\global\advance\figno by 1
\midinsert
\epsfxsize=#3
\centerline{\epsfbox{#2}}
\vskip 12pt
{\bf Fig. \the\figno:~~} #1 \par
\endinsert\endgroup\par
}
\def\figlabel#1{\xdef#1{\the\figno}}
\newdimen\tableauside\tableauside=1.0ex
\newdimen\tableaurule\tableaurule=0.4pt
\newdimen\tableaustep
\def\phantomhrule#1{\hbox{\vbox to0pt{\hrule height\tableaurule
width#1\vss}}}
\def\phantomvrule#1{\vbox{\hbox to0pt{\vrule width\tableaurule
height#1\hss}}}
\def\sqr{\vbox{%
  \phantomhrule\tableaustep

\hbox{\phantomvrule\tableaustep\kern\tableaustep\phantomvrule\tableaustep}%
  \hbox{\vbox{\phantomhrule\tableauside}\kern-\tableaurule}}}
\def\squares#1{\hbox{\count0=#1\noindent\loop\sqr
  \advance\count0 by-1 \ifnum\count0>0\repeat}}
\def\tableau#1{\vcenter{\offinterlineskip
  \tableaustep=\tableauside\advance\tableaustep by-\tableaurule
  \kern\normallineskip\hbox
    {\kern\normallineskip\vbox
      {\gettableau#1 0 }%
     \kern\normallineskip\kern\tableaurule}%
  \kern\normallineskip\kern\tableaurule}}
\def\gettableau#1 {\ifnum#1=0\let\next=\null\else
  \squares{#1}\let\next=\gettableau\fi\next}

\tableauside=1.0ex
\tableaurule=0.4pt


\def\Q{{\cal Q}}

\def\th{\theta}
\def\Th{\Theta}

\def\hf{{1\over 2}}

\def\o{\over}

\def\til#1{\widetilde{#1}}
\def\si{\sigma}
\def\Si{\Sigma}

\def\del{\partial}

\def\lf{\left}
\def\ri{\right}
\def\riya{\rightarrow}

\def\h#1{\widehat{#1}}

\def\bt{\beta}
\def\ga{\gamma}
\def\Ga{\Gamma}

\def\om{\omega}

\def\dag{\dagger}

\def\sitarel#1#2{\mathrel{\mathop{\kern0pt #1}\limits_{#2}}}

\lref\AharonyUG{
  O.~Aharony, O.~Bergman, D.~L.~Jafferis and J.~Maldacena,
  ``N=6 superconformal Chern-Simons-matter theories, M2-branes and their
  gravity duals,''
  arXiv:0806.1218 [hep-th].
}
\lref\BennaZY{
  M.~Benna, I.~Klebanov, T.~Klose and M.~Smedback,
  ``Superconformal Chern-Simons Theories and AdS$_4$/CFT$_3$ Correspondence,''
  arXiv:0806.1519 [hep-th].
}
\lref\ImamuraNN{
  Y.~Imamura and K.~Kimura,
  ``Coulomb branch of generalized ABJM models,''
  arXiv:0806.3727 [hep-th].
}
\lref\HananyQC{
  A.~Hanany, N.~Mekareeya and A.~Zaffaroni,
  ``Partition Functions for Membrane Theories,''
  arXiv:0806.4212 [hep-th].
}
\lref\GauntlettPK{
  J.~P.~Gauntlett, G.~W.~Gibbons, G.~Papadopoulos and P.~K.~Townsend,
  ``Hyper-Kaehler manifolds and multiply intersecting branes,''
  Nucl.\ Phys.\  B {\bf 500}, 133 (1997)
  [arXiv:hep-th/9702202].
}
\lref\TongRQ{
  D.~Tong,
  ``NS5-branes, T-duality and worldsheet instantons,''
  JHEP {\bf 0207}, 013 (2002)
  [arXiv:hep-th/0204186].
}
\lref\GibbonsNT{
  G.~W.~Gibbons and P.~Rychenkova,
  ``HyperKaehler quotient construction of BPS monopole moduli spaces,''
  Commun.\ Math.\ Phys.\  {\bf 186}, 585 (1997)
  [arXiv:hep-th/9608085].
}
\lref\RocekPS{
  M.~Rocek and E.~P.~Verlinde,
  ``Duality, quotients, and currents,''
  Nucl.\ Phys.\  B {\bf 373}, 630 (1992)
  [arXiv:hep-th/9110053].
}
\lref\HarveyAB{
  J.~A.~Harvey and S.~Jensen,
  ``Worldsheet instanton corrections to the Kaluza-Klein monopole,''
  JHEP {\bf 0510}, 028 (2005)
  [arXiv:hep-th/0507204].
}
\lref\WittenYC{
  E.~Witten,
  ``Phases of N = 2 theories in two dimensions,''
  Nucl.\ Phys.\  B {\bf 403}, 159 (1993)
  [arXiv:hep-th/9301042].
}
\lref\ColemanCI{
  S.~R.~Coleman,
  ``There are no Goldstone bosons in two-dimensions,''
  Commun.\ Math.\ Phys.\  {\bf 31}, 259 (1973).
}
\lref\OkuyamaGX{
  K.~Okuyama,
  ``Linear sigma models of H and KK monopoles,''
  JHEP {\bf 0508}, 089 (2005)
  [arXiv:hep-th/0508097].
}
\lref\HawkingJB{
  S.~W.~Hawking,
  ``Gravitational Instantons,''
  Phys.\ Lett.\  A {\bf 60}, 81 (1977).
}
\lref\GibbonsXM{
  G.~W.~Gibbons and S.~W.~Hawking,
  ``Classification Of Gravitational Instanton Symmetries,''
  Commun.\ Math.\ Phys.\  {\bf 66}, 291 (1979).
}
\lref\Kronheimer{
 P.~Kronheimer, 
``The construction of ALE spaces as hyper-kahler quotients,''
J. Diff. Geom. {\bf 28}, 665 (1989);
``A Torelli-Type Theorem for Gravitational Instantons,''
J. Diff. Geom. {\bf 29}, 685 (1989).
}
\lref\BaggerVI{
  J.~Bagger and N.~Lambert,
  ``Comments On Multiple M2-branes,''
  JHEP {\bf 0802}, 105 (2008)
  [arXiv:0712.3738 [hep-th]].
}
\lref\BaggerJR{
  J.~Bagger and N.~Lambert,
  ``Gauge Symmetry and Supersymmetry of Multiple M2-Branes,''
  Phys.\ Rev.\  D {\bf 77}, 065008 (2008)
  [arXiv:0711.0955 [hep-th]].
}
\lref\BaggerSK{
  J.~Bagger and N.~Lambert,
  ``Modeling multiple M2's,''
  Phys.\ Rev.\  D {\bf 75}, 045020 (2007)
  [arXiv:hep-th/0611108].
}
\lref\GustavssonVU{
  A.~Gustavsson,
  ``Algebraic structures on parallel M2-branes,''
  arXiv:0709.1260 [hep-th].
}
\lref\EguchiXP{
  T.~Eguchi and A.~J.~Hanson,
  ``Asymptotically Flat Selfdual Solutions To Euclidean Gravity,''
  Phys.\ Lett.\  B {\bf 74}, 249 (1978).
}

\Title{             
}
{\vbox{
\centerline{Linear Sigma Models for the $R^8/{\cal Z}_k$ Orbifold}
}}

\vskip .2in

\centerline{Kazumi Okuyama}
\vskip5mm
\centerline{Department of Physics, Shinshu University}
\centerline{Matsumoto 390-8621, Japan}
\centerline{\tt kazumi@azusa.shinshu-u.ac.jp}
\vskip .2in

\vskip 3cm
\noindent

We construct ${\cal N}=4$ gauged linear sigma
models in two dimensions whose Higgs branch has a 
${\Bbb R}^8/{\Bbb Z}_k$ orbifold singularity or its generalization. 
Our linear sigma models have either
ALF or ALE type hyperK\"{a}hler 8-manifolds as their
Higgs branch. For the ALE case, the matter content of our model
is specified by a quiver diagram which is
a union of two $A$-type extended Dynkin diagrams overlapping
at one link.

\Date{July 2008}
\vfill
\vfill

\newsec{Introduction}
It is well-known that some hyperK\"{a}hler manifolds, such as 
Taub-NUT \refs{\HawkingJB,\GibbonsXM} and ALE spaces \EguchiXP,
can be realized as hyperK\"{a}hler quotients of flat ${\Bbb R}^{4k}$ 
\refs{\Kronheimer,\GibbonsNT}.
In this paper, following the general procedure
in \GauntlettPK, we will consider the quotient construction of
some 8-dimensional hyperK\"{a}hler manifolds which have
${\Bbb R}^8/{\Bbb Z}_k$ orbifold singularity in 
a certain limit. 

This is partly motivated by the recent 
excitement of the Bagger-Lambert-Gustavsson
theory of multiple M2-branes 
\refs{\BaggerVI,\BaggerJR,\BaggerSK,\GustavssonVU}
and the closely related model by Aharony, Bergman, Jafferis and Maldacena
(ABJM) \AharonyUG.
For the ABJM model, {\it i.e.} a three-dimensional ${\cal N}=6$ $U(N)\times U(N)$
Chern-Simons-matter theory with level $(k,-k)$ \AharonyUG, it is shown  that
its vacuum moduli space is $({\Bbb R}^8/{\Bbb Z}_k)^N/S_N$, 
which suggests that this model is a theory on
M2-branes in the orbifold ${\Bbb R}^8/{\Bbb Z}_k$ background.
As argued in \AharonyUG, 
this picture is also consistent with the brane construction of
the model. Namely, the ABJM model 
is realized as a theory on the D3-branes wrapped around a 
circle in the presence of a NS5-brane and a $(k,1)$ 5-brane
transverse to the circle.
The M-theory dual of this configuration is a collection of
M2-branes in the background of intersecting KK monopoles.
The corresponding 
11-dimensional supergravity solution 
is given by an 8-dimensional toric hyperK\"{a}hler manifold \GauntlettPK.
It is shown that the hyperK\"{a}hler manifold appearing as the
dual of NS5-$(k,1)$5brane system has a ${\Bbb R}^8/{\Bbb Z}_k$ orbifold
singularity \AharonyUG. Some generalizations
of the ABJM model, which correspond to more general orbifold
${\Bbb R}^8/\Ga$, were considered in \refs{\BennaZY,\ImamuraNN}.

In this paper, we 
will construct two-dimensional ${\cal N}=4$ gauged linear
sigma models (GLSMs) whose Higgs branch is a hyperK\"{a}hler manifold
which appears as the M-theory dual of a configuration of
$n$ NS5-branes and $k$ $(1,1)$ 5-branes, or
$n$ NS5-branes and one $(k,1)$ 5-brane.
We should emphasize that our GLSM is {\it not}
directly related to the theory on M2-branes in the orbifold
background. We merely use GLSM as a tool to
realize the hyperK\"{a}hler quotient construction
in the gauge theory language.
Our GLSM is a natural generalization of the 
model for the Taub-NUT space 
studied in \refs{\HarveyAB,\OkuyamaGX}, which was shown to be dual to
the GLSM for H-monopoles \refs{\TongRQ,\OkuyamaGX} applying 
the method of \RocekPS.
We consider both ALF and ALE type hyperK\"{a}hler 8-manifolds,
presented in section 2 and 3, respectively. 
For the ALE case, the matter content of our GLSM is
described by a quiver diagram, which is a union of $\h{A}_{k-1}$ and 
$\h{A}_{n-1}$ Dynkin diagrams connected at one link (see Fig. 1).
 
\newsec{ALF-type GLSM}
We first construct an ${\cal N}=4$ GLSM in two dimensions 
whose Higgs branch is an ALF-type
hyperK\"{a}hler 8-manifold, which appears as
the M-theory dual of the type IIB 5-brane configurations. 
In the case of H-monopoles or its T-dual of KK-monopoles,
the corresponding GLSMs were 
studied in \refs{\TongRQ,\HarveyAB,\OkuyamaGX}.
Let us recall the matter content of
the GLSM for the Taub-NUT space with KK-monopole charge $k$ 
\refs{\GibbonsNT,\OkuyamaGX}.
The model has the gauge group $\prod_{a=1}^kU(1)_a$
with $k$ hypermultiplets $(Q_a,\til{Q}_a)$ with charge $(+1,-1)$ 
under the gauge
group $U(1)_a$. Additionally, there is a linear-multiplet $(\Psi,P)$,
where the shift symmetry of the imaginary part of $P$ is gauged
under the diagonal part of $\prod_aU(1)_a$. 

\subsec{M-theory Dual of $n$ NS5-branes and $k$ $(1,1)$ 5-branes}
We first consider the GLSM for the 8-manifold which is dual to
a configuration of $n$ NS5-branes and $k$ $(1,1)$ 5-branes.
Since this brane configuration of  
5-branes is U-dual to the intersecting KK-monopoles,
we expect that the GLSM for this background is
obtained by a simple generalization of
the Taub-NUT case. 
We will show that this is indeed the case
followng the general recipe for the
quotient construction of
toric hyperK\"{a}hler 8-manifolds \GauntlettPK.
The matter content of our GLSM is the same
as the two sets of GLSMs for Taub-NUT spaces with charge $k$ and
$n$, which we call $A$-part and $B$-part, respectively:
\eqn\mattercont{\eqalign{
&A{\rm -part}\quad\lf\{\eqalign{
{\rm vector}:\quad & (\Si_a,\Phi_a)\cr
{\rm hyper}:\quad & (Q_a,\til{Q}_a)\cr
{\rm linear}:\quad & (\Psi_A,P_A)
}\ri.\qquad (a=1,\cdots,k) \cr
&B{\rm -part}\quad\lf\{\eqalign{
{\rm vector}:\quad & (\Si_i,\Phi_i)\cr
{\rm hyper}:\quad & (H_i,\til{H}_i)\cr
{\rm linear}:\quad & (\Psi_B,P_B)~~.
}\ri.\qquad (i=1,\cdots,n)
}}
Here and in what follows,
we use the ${\cal N}=2$ language as in \WittenYC. For instance,
$\Si_a$ and $\Si_i$ are the twisted chiral multiplets in the ${\cal N}=2$
language. All other fields such as $\Q_a$ and $\Phi_a$
are ${\cal N}=2$ chiral multiplets.
The gauge groups of the $A$-part and the $B$-part are
$\prod_{a=1}^kU(1)_{A,a}$ and $\prod_{i=1}^nU(1)_{B,i}$.
The only difference from the naive direct sum of two Taub-NUT models is
that the linear-multiplet in the $B$-part is shifted by
the diagonal part of the total gauge group
$\prod_{a=1}^kU(1)_{A,a}\times\prod_{i=1}^nU(1)_{B,i}$,
while the linear-multiplet of the $A$-part is shifted only by
the diagonal of $\prod_{a=1}^kU(1)_{A,a}$ 
as in the original Taub-NUT model.

The Lagrangian of our model \mattercont\
is given by ${\cal L}={\cal L}_D+{\cal L}_F+
{\cal L}_{\tilde{F}}$, where the D-term ${\cal L}_D$ is
\eqn\lagD{\eqalign{
{\cal L}_D&=\int d^4\th~{1\o g_A^2}\Psi_A^\dag\Psi_A+{g_A^2\o2}
\lf(P_A+P_A^\dag+\sum_{a=1}^kV_a\ri)^2\cr
&\hskip10mm +{1\o g_B^2}\Psi_B^\dag\Psi_B+{g_B^2\o2}
\lf(P_B+P_B^\dag+\sum_{a=1}^kV_a+\sum_{i=1}^nV_i\ri)^2\cr
&\hskip5mm+\sum_{a=1}^k\lf\{{1\o e_a^2}(-\Si_a^\dag\Si_a+\Phi_a^\dag\Phi_a)
+Q_a^\dag e^{V_a}Q_a+\til{Q}_a^\dag e^{-V_a}\til{Q}_a\ri\}\cr
&\hskip5mm+\sum_{i=1}^n\lf\{{1\o e_i^2}(-\Si_i^\dag\Si_i+\Phi_i^\dag\Phi_i)
+H_i^\dag e^{V_i}H_i+\til{H}_i^\dag e^{-V_i}\til{H}_i\ri\}~,
}}
and the F-term ${\cal L}_F$ and the twisted F-term ${\cal L}_{\tilde{F}}$
are
\eqn\lagF{\eqalign{
{\cal L}_F&=\int d\th^+ d\th^-\sum_{a=1}^k\lf\{\til{Q}_a\Phi_aQ_a+
(s_a-\Psi_A)\Phi_a\ri\}+
\sum_{i=1}^n\lf\{\til{H}_i\Phi_iH_i+
(s_i-\Psi_B)\Phi_i\ri\}+c.c.\cr
{\cal L}_{\tilde{F}}&=\int d\th^+ d\bar{\th}^-\sum_{a=1}^kt_a\Si_a+\sum_{i=1}^n
t_i\Si_i+c.c.~.
}}
In the above equations,
$e_a^2$ and $e_i^2$ denote the gauge couplings, and $g_A^2$
and $g_B^2$ are some parameters.
The parameters $(s_a,t_a)$ and $(s_i,t_i)$ appearing in \lagF\
are the ${\cal N}=4$ FI-parameters.
They are naturally decomposed into the triplets $(\vec{r}_a,\vec{r}_i)$
and the singlets $(\th_a,\th_i)$ under the $SU(2)_R$ R-symmetry:
\eqn\FIst{
s_a=r_a^1+ir_a^2,\quad t_a=r_a^3+i\th_a,\quad
s_i=r_i^1+ir_i^2,\quad t_i=r_i^3+i\th_i.
}
In terms of the component fields, the bosonic part
our Lagrangian is written as a sum of 
the kinetic term ${\cal L}_{\rm kin}$, the potential term ${\cal L}_{\rm pot}$
and the topological term ${\cal L}_{\rm top}$:
\eqn\Lagboskin{\eqalign{
{\cal L}_{\rm kin}=&{1\o2g_A^2}(\del \vec{x}_A)^2+{g_A^2\o2}
\Big(\del\ga_A+\sum_{a=1}^kA_a\Big)^2\cr
+&{1\o2g_B^2}(\del \vec{x}_B)^2+{g_B^2\o2}
\Big(\del\ga_B+\sum_{a=1}^kA_a+\sum_{i=1}^nB_i\Big)^2\cr
+&\sum_{a=1}^k\lf\{{1\o e_a^2}\Big((F_{01}^a)^2+
|\del\phi_a|^2+|\del\si_a|^2\Big)+|{\cal D}q_a|^2+|{\cal D}\til{q}_a|^2\ri\}
\cr
+&\sum_{i=1}^n\lf\{{1\o e_i^2}\Big((F_{01}^i)^2+
|\del\phi_i|^2+|\del\si_i|^2\Big)+|{\cal D}h_i|^2+|{\cal D}\til{h}_i|^2\ri\}
}}
\eqn\lagpot{\eqalign{
{\cal L}_{\rm pot}=&-\sum_{a=1}^k\Big\{
{e_a^2\o2}\big(|q_a|^2-|\til{q}_a|^2-x_A^3-x_B^3+r_a^3\big)^2\cr
&\hskip8mm
+{e_a^2\o2}\big|2q_a\til{q}_a-(x_A^1+x_B^1+ix_A^2+ix_B^2)+r_a^1+ir_a^2\big|^2
\cr
&\hskip12mm+\big(|\phi_a|^2+|\si_a|^2\big)
\big(|q_a|^2+|\til{q}_a|^2+g_A^2\big)
\Big\}\cr
&-\sum_{i=1}^n\Big\{
{e_i^2\o2}\big(|h_i|^2-|\til{h}_i|^2-x_B^3+r_i^3\big)^2\cr
&\hskip8mm
+{e_i^2\o2}\big|2h_i\til{h}_i-(x_B^1+ix_B^2)+r_i^1+ir_i^2\big|^2
\cr
&\hskip12mm+\big(|\phi_i|^2+|\si_i|^2\big)\big(|h_i|^2+|\til{h}_i|^2+g_B^2\big)
\Big\}
}}
\eqn\lagtop{
{\cal L}_{\rm top}=-\sum_{a=1}^k\th_aF_{01}^a-
\sum_{i=1}^n\th_iF_{01}^i.
}
Here we used the lower case letters to denote
the scalar components of the corresponding (twisted) chiral superfields,
except for the linear-multiplets.
For the linear-multiplets, the scalar components
are denoted as
\eqn\scalPAB{
\Psi_A=x_A^1+ix_A^2,\quad \Psi_B=x_B^1+ix_B^2,\quad
P_A={1\o g_A^2}x_A^3+i\ga_A,\quad 
P_B={1\o g_B^2}x_B^3+i\ga_B.
}
$A_a=A_{a,\mu}dx^\mu$ and $B_i=B_{i,\mu}dx^\mu$
in \Lagboskin\ are the gauge fields for the gauge groups
$U(1)_{A,a}$ and $U(1)_{B,i}$, respectively.
$\vec{x}_A$ and $\vec{x}_B$ appearing in \Lagboskin\
denote the $SU(2)_R$ triplet parts of the scalar components
of the linear-multiplets \scalPAB
\eqn\xvec{
\vec{x}_A=(x_A^1,x_A^2,x_A^3),\quad
\vec{x}_B=(x_B^1,x_B^2,x_B^3).
}
The kinetic term in \Lagboskin\ such as $(\del\vec{x}_A)^2$ means
$\sum_{\mu=0,1}\del_\mu\vec{x}_A\cdot\del^\mu\vec{x}_A$.
$\ga_A$ and $\ga_B$ are normalized to have the period $2\pi$
\eqn\gaperiod{
\ga_A\sim\ga_A+2\pi,\quad
\ga_B\sim\ga_B+2\pi~.
}

In the rest of this section, we will analyze
the Higgs branch of our model. From the expression of the
potential energy in \lagpot, the vacuum moduli space\foot{
Strictly speaking, there is no moduli space of vacua in two dimensions
because of the Coleman theorem \ColemanCI. We analyze the
low energy theory in the spirit of
Born-Oppenheimer approximation.
}
is characterized by
\eqn\vaceq{\eqalign{
&F_{01}^a=F_{01}^i=\si_a=\phi_a=\si_i=\phi_i=0\cr
&|q_a|^2-|\til{q}_a|^2=x_A^3+x_B^3-r_a^3,\quad
2q_a\til{q}_a=x_A^1+x_B^1+i(x_A^2+x^2_B)-r_a^1-ir_a^2\cr
&|h_i|^2-|\til{h}_i|^2=x_B^3-r_i^3,\quad
2h_i\til{h}_i=x_B^1+ix^2_B-r_i^1-ir_i^2.
}}
In the IR limit $e_a^2,e_i^2\riya\infty$, the vector multiplets and the 
charged hypermultiplets become massive
and they can be integrated out.
To find the low energy action,
the crucial step is to rewrite
the kinetic term of hypermultiplet restricted on
the vacuum locus \vaceq
\eqn\kinhyp{\eqalign{
|{\cal D}q_a|^2+|{\cal D}\til{q}_a|^2&=
{\big(\del\vec{x}_A+\del\vec{x}_B\big)^2\o4|\vec{x}_A+\vec{x}_B-\vec{r}_a|}
+{|\vec{x}_A+\vec{x}_B-\vec{r}_a|\o4}\Big\{2A_a+2\del\varphi_a+\vec{\om}_a\cdot
(\del\vec{x}_A+\del\vec{x}_B)\Big\}^2\cr
|{\cal D}h_i|^2+|{\cal D}\til{h}_i|^2&=
{(\del\vec{x}_B)^2\o4|\vec{x}_B-\vec{r}_i|}
+{|\vec{x}_B-\vec{r}_i|\o4}\Big(2B_i+2\del\varphi_i+\vec{\tau}_i\cdot
\del\vec{x}_B\Big)^2~,
}}
where $\varphi_a=-{\rm arg}(iq_a)$ and $\varphi_i=-{\rm arg}(ih_i)$. 
$\vec{\om}_a$ and $\vec{\tau}_i$ in the above equations are given by
\eqn\omtaunab{
\vec{\nabla}\times\vec{\om}_a=\vec{\nabla}{1\o|\vec{x}_A+\vec{x}_B-\vec{r}_a|},\quad
\vec{\nabla}\times\vec{\tau}_i=\vec{\nabla}{1\o|\vec{x}_B-\vec{r}_i|}.
}
Due to the gauge symmetry, the low energy theory
depends only on the gauge invariant combinations  
\eqn\thABinv{
\th_A=\ga_A-\sum_{a=1}^k\varphi_a,\quad
\th_B=\ga_B-\sum_{a=1}^k\varphi_a-\sum_{i=1}^n\varphi_i.
}
In the IR limit the gauge kinetic term can be ignored, hence the gauge
fields $A_a$ and $B_i$ become auxiliary fields.
After integrating out the gauge fields, we arrive at
the effective Lagrangian on the Higgs branch
\eqn\Leffmet{
{\cal L}_{\rm eff}=\hf\sum_{i,j=A,B}\lf(U_{ij}\del\vec{x}_i\cdot\del\vec{x}_j+
(U^{-1})_{ij}\bt_i\bt_j\ri)
}
where $\bt_A$ and $\bt_B$ are given by
\eqn\xonetwo{
\bt_A=\del\th_A
-\hf\sum_{a=1}^k\vec{\om}_a\cdot(\del\vec{x}_A+\del\vec{x}_B),\quad
\bt_B=\del\th_B
-\hf\sum_{a=1}^k\vec{\om}_a\cdot(\del\vec{x}_A+\del\vec{x}_B)-\hf\sum_{i=1}^n
\vec{\tau}\cdot\del\vec{x}_B,
}
and the matrix $U$ in \Leffmet\ is
\eqn\Umat{\eqalign{
U&=\lf(\matrix{U_{AA}&U_{AB}\cr U_{BA}&U_{BB}}\ri)
=\lf(\matrix{{1\o g_A^2}+H&H\cr H&{1\o g_B^2}+K+H}\ri), \cr
H&=\hf\sum_{a=1}^k{1\o|\vec{x}_A+\vec{x}_B-\vec{r}_a|},\quad
K=\hf\sum_{i=1}^n{1\o|\vec{x}_B-\vec{r}_i|}.
}}
For the $n=k=1$ case, 
one can easily see that the effective metric on the Higgs branch
is nothing but the metric studied in \AharonyUG,
which was shown to be the M-theory dual of
a NS5-brane and a $(1,1)$ 5-brane. 
For the general  case,
the metric becomes
singular  when $K\riya\infty$ or $H\riya\infty$.
This implies that
when we set $\vec{r}_a=\vec{r}_i=0$
there is a singularity at the origin $\vec{x}_A=\vec{x}_B=0$.
Near the origin the metric behaves as
\eqn\Lefforigin{
{\cal L}_{\rm eff}\sim\hf\Big\{H(\del\vec{x}_A+\del\vec{x}_B)^2+
K(\del\vec{x}_B)^2+H^{-1}(\bt_A)^2+K^{-1}(\bt_A-\bt_B)^2\Big\}.
}
From this expression, one can see that the moduli space has a
${\Bbb R}^4/{\Bbb Z}_k\times{\Bbb R}^4/{\Bbb Z}_n$ orbifold singularity. 
From the constant part $U_\infty$ of the matrix $U$
\eqn\Uinfinity{
U_{\infty}=\lf(\matrix{{1\o g_A^2}&0\cr0&{1\o g_B^2}}\ri),
}
we can read off the moduli of the torus (or type IIB axio-dilaton) as  
\AharonyUG
\eqn\IIBtau{
\tau=\chi+{i\o g_s}=i{g_B\o g_A}.
}
The singularity at the origin
is factorized ${\Bbb R}^4/{\Bbb Z}_k\times{\Bbb R}^4/{\Bbb Z}_n$
since the configuration with $n$ NS5-brane and $k$ $(1,1)$ 5-branes
beomes equivalent to the configuration
of $n$ NS5-brane and $k$ D5-branes
by the shift $\tau\riya\tau+1$.
The latter configuration is
dual to the orthogonal KK-monopoles, hence
the singularity is factorized.

As discussed in \refs{\HarveyAB,\OkuyamaGX},
we can perform T-duality along one of the $S^1$ direction, say $\th_B$,
by using the method of \RocekPS.
In this duality, the linear-multiplet $(\Psi_B,P_B)$
is replaced by the twisted hypermultiplet $(\Psi_B,\Th)$
where $\Th$ is a twisted chiral multiplet in the ${\cal N}=2$
language. The resulting model describes the 
configuration of $n$ NS5-branes intersecting with KK-monopoles.
As argued in \refs{\TongRQ,\HarveyAB,\OkuyamaGX},
the low energy effective action
receives instanton corrections, which leads to
the localization
of brane positions along the $S^1$ direction. 
It would be interesting to study such instanton corrections in
our model.

\subsec{M-theory Dual of $n$ NS5-branes and one $(k,1)$ 5-brane}
Next we consider the the GLSM for the configuration of
$n$ NS5-branes and one $(k,1)$ 5-brane. This is obtained
by replacing the $A$-part in the previous subsection
with the following model of single $U(1)_A$ gauge symmetry: 
one hypermulriplet
with charge $1$ under the gauge group $U(1)_A$,  
and the 
linear-multiplet with shift charge $k$  
under $U(1)_A$.
The linear multiplet in the $B$-part is charged under 
the diagonal of $U(1)_A\times\prod_{i=1}^nU(1)_{B,i}$.
The D-term for the linear multiplet reads
\eqn\LDforkone{\eqalign{
{\cal L}_{D}^{\rm linear}&=
\int d^4\th~{1\o g_A^2}\Psi_A^\dag\Psi_A+{g_A^2\o2}
\lf(P_A+P_A^\dag+kV_A\ri)^2\cr
&\hskip10mm +{1\o g_B^2}\Psi_B^\dag\Psi_B+{g_B^2\o2}
\lf(P_B+P_B^\dag+V_A+\sum_{i=1}^nV_i\ri)^2
}}
where $V_A$ is the vector superfield for the gauge group $U(1)_A$.
After a similar analysis as in the previous subsection,
we find that the effective metric on the Higgs branch has
the same form as \Lefforigin\ with 
\eqn\Ubtkone{\eqalign{
U&=\lf(\matrix{{1\o g_A^2}+k^2H&kH\cr
kH&{1\o g_B^2}+K+H}\ri), \quad
H={1\o2|k\vec{x}_A+\vec{x}_B|},\quad 
K=\hf\sum_{i=1}^n{1\o|\vec{x}_B-\vec{r}_i|}\cr
\bt_A&=\del\th_A-{k\o2}\vec{\om}\cdot(k\del\vec{x}_A+\del\vec{x}_B),
\quad \bt_B=\del\th_B-{1\o2}\vec{\om}\cdot(k\del\vec{x}_A+\del\vec{x}_B)
-\hf\sum_{i=1}^n\vec{\tau}_i\cdot\del\vec{x}_B.
}
}
By the similar analysis as in \AharonyUG,
we find that 
the metric has the orbifold singularity ${\Bbb C}^4/\Ga$,
where $\Ga$ is generated by $g_1$ and $g_2$
\eqn\gotwo{\eqalign{
g_1:\quad & (z_1,z_2,z_3,z_4)\sim (e^{2\pi i\o k}z_1,e^{-{2\pi i\o k}}z_2,
e^{2\pi i\o kn}z_3,e^{-{2\pi i\o kn}}z_4)\cr
g_2:\quad & (z_1,z_2,z_3,z_4)\sim (z_1,z_2,
e^{2\pi i\o n}z_3,e^{-{2\pi i\o n}}z_4).
}}
In particular, the singularity for the $n=1$ case is
${\Bbb R}^8/{\Bbb Z}_k$ \AharonyUG. 
For the general case, \gotwo\ is in agreement with \ImamuraNN.

\newsec{ALE-type GLSM (or Quiver Gauge Theory)}
In this section, we will consider the ALE analogue of the model.
The ALE-type GLSM can be obtained from the ALF-type cousin studied
in section 2.1. Let us first consider the $A$-part.
We replace the hypermultiplet $(Q_a,\til{Q}_a)$
charged under $U(1)_{A,a}$ by the ``bi-fundamental'' hypermultiplet
charged under $U(1)_{A,a}\times U(1)_{A,a+1}$.
In order to have the $A_{k-1}$ model, we have to reduce the
number of hypermultiplets by one, {\it i.e.} $a$ runs from $2$ to $k$.
We should also
promote the linear-multiplet to a ``bi-fundamental'' hypermultiplet
charged under $U(1)_{A,1}\times U(1)_{A,2}$.
Then the gauge field appearing the Lagrangian \Lagboskin\
is replaced as
\eqn\Areplace{\eqalign{
&A_a\riya A_a-A_{a+1}\cr
&\sum_{a=1}^{k}A_a\riya \sum_{a=2}^{k} (A_a-A_{a+1})=A_2-A_1
}} 
where we identified $k+1\equiv 1$.
Then the resulting theory is described by the $\h{A}_{k-1}$ Dynkin diagram.
Note that the link between the node 1 and node 2 represents the
hypermultiplet coming from the linear-multiplet in the ALF-type model
in the previous section.

We can do the same replacement in the $B$-part.
Then we get a matter content specified by the $\h{A}_{n-1}$
Dynkin diagram. However, there is an important difference for the
link between the node 1 and node 2 from the rest of the links. Since the
linear-multiplet for the $B$-part is charged under the
gauge field $\sum_aA_a+\sum_iB_i$ for the ALF case, 
this becomes a hypermultiplet in the ALE model
charged under the gauge field
\eqn\onetwoAB{
\sum_{a=1}^kA_a+\sum_{i=1}^nB_i\riya
\sum_{a=2}^k(A_a-A_{a+1})+\sum_{i=2}^n(B_i-B_{i+1})
=A_2-A_1+B_2-B_1.
}
Therefore, the hypermultiplet
on the link between the node 1 and node 2 in the
$B$-part 
is charged under $U(1)_{A,1}\times U(1)_{A,2}\times U(1)_{B,1}
\times U(1)_{B,2}$.

\fig{The quiver diagram for the ALE-type GLSM is a union of 
the $\h{A}_{k-1}$ Dynkin diagram (labeled $A$, black)
and the $\h{A}_{n-1}$ Dynkin diagram (labeled $B$, blue).
The link between the node 1 and 2
in the diagram $B$ 
(the dashed line between the node 1 and 2)
is charged under $U(1)_{A,1}\times U(1)_{A,2}\times 
U(1)_{B,1}\times U(1)_{B,2}$, while
the link between the node 1 and 2 
in the  diagram $A$ (the solid line between the node 1 and 2),
is charged only under $U(1)_{A,1}\times U(1)_{A,2}$.
}{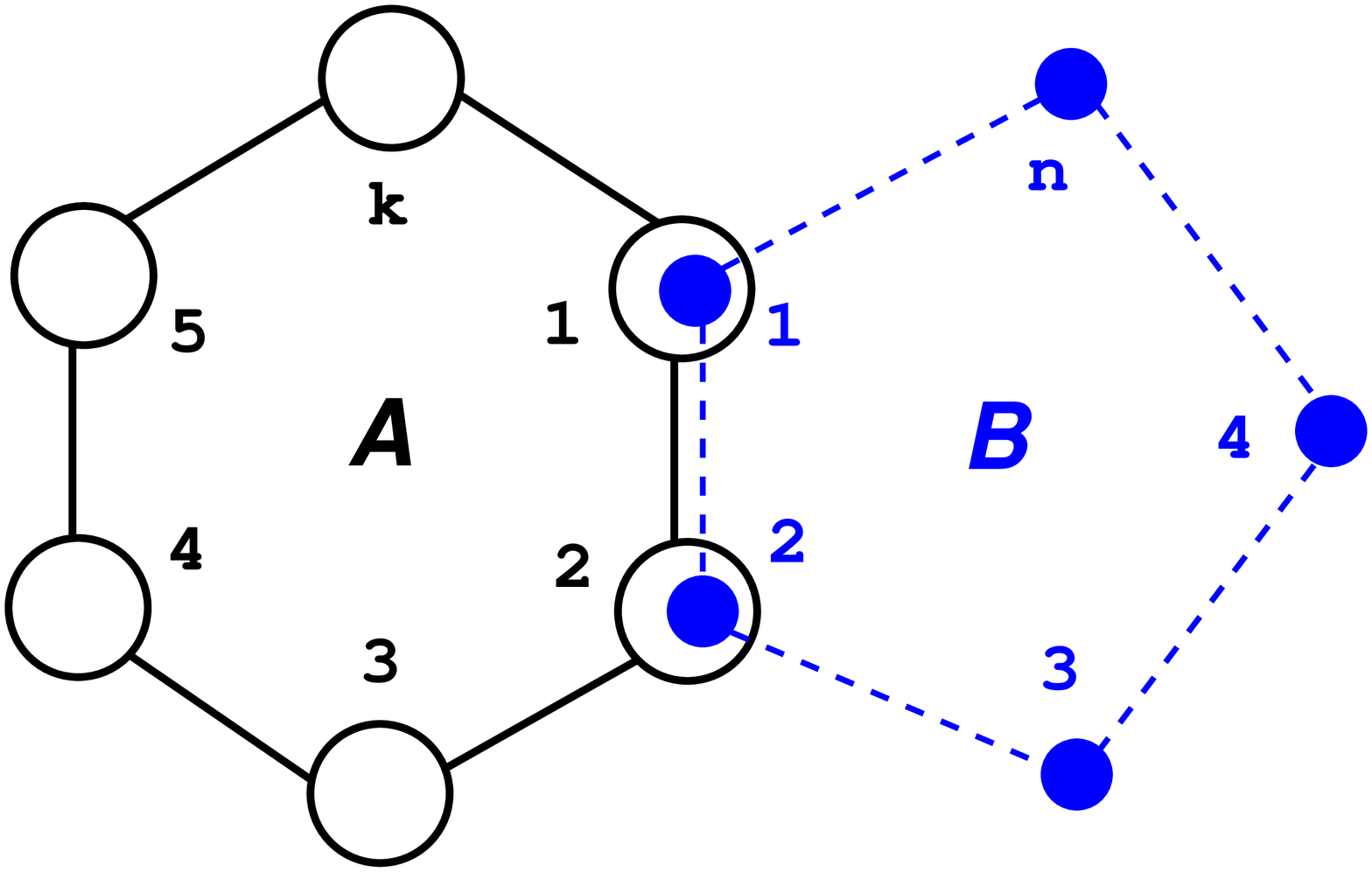}{9cm}

The resulting matter content of the ALE-type GLSM is summarized 
by the quiver diagram in Fig. 1.
Namely, the quiver diagram of our theory
is a union of two $\h{A}$ Dynkin diagrams overlapping at the link between
the node 1 and node 2. The only difference from the usual
ALE quiver is that the link between the node 1 and 2 in the diagram $B$
is charged under both $U(1)_{B,1}\times U(1)_{B,2}$ and
$U(1)_{A,1}\times U(1)_{A,2}$.
Other links in the diagram $A$ (resp. diagram $B$)
are charged only under the gauge group $U(1)_{A,a}\times U(1)_{A,a+1}$
(resp. $U(1)_{B,i}\times U(1)_{B,i+1}$).

\subsec{Singularity of the Higgs branch}
Now we consider the singularity of the moduli space.
It is straightforward to study the low energy effective
metric on the Higgs branch as in the
previous section. The resulting metric
is not the one obtained form the ALF case \Leffmet\
by setting the constant part $U_\infty$ of the
matrix $U$ to zero.
Instead of analyzing the metric,
let us consider the singularity
from the complex viewpoint by looking at the
F-term constraints for the Higgs branch: 
\eqn\Ftermeq{\eqalign{
& q_{a,a+1}\til{q}_{a+1,a}-q_{a-1,a}\til{q}_{a,a-1}=\mu_a
\quad(a=3,\cdots,k)\cr
&q_{1,2}\til{q}_{2,1}-q_{k,1}\til{q}_{1,k}+h_{1,2}\til{h}_{2,1}=\mu_1\cr
&q_{2,3}\til{q}_{3,2}-q_{1,2}\til{q}_{2,1}-h_{1,2}\til{h}_{2,1}=\mu_2\cr
&h_{i,i+1}\til{h}_{i+1,i}-h_{i-1,i}\til{h}_{i,i-1}=\zeta_i
\quad(i=1,\cdots n)~,
}}
where $\mu_a$ and $\zeta_i$ are the complex FI-parameters.
For the consistency of these relations, the FI-parameters should
satisfy
\eqn\summuz{
\sum_{a=1}^k\mu_a=\sum_{i=1}^n\zeta_i=0.
}
Then the equations \Ftermeq\ can be solved as
\eqn\qhinuv{\eqalign{
&q_{1,2}\til{q}_{2,1}=u-v,~\qquad h_{1,2}\til{h}_{2,1}=v~,\cr
&q_{a,a+1}\til{q}_{a+1,a}=u+c_a~;\quad c_a=\sum_{b=2}^a\mu_b~, 
\quad(a=2,\cdots k)~,\cr
&h_{i,i+1}\til{h}_{i+1,i}=v+d_i~;\quad d_i=\sum_{j=2}^i\zeta_j~, 
\qquad(i=2,\cdots n)~.
}} 
By introducing the baryonic operators
\eqn\xyzwbary{\eqalign{
&x=q_{1,2}q_{2,3}\cdots q_{k,1}\cr
&y=\til{q}_{2,1}\til{q}_{3,2}\cdots\til{q}_{1,k}\cr
&z=h_{1,2}h_{2,3}\cdots h_{n,1}\cr
&w=\til{h}_{2,1}\til{h}_{3,2}\cdots\til{h}_{1,n}
}}
the vacuum moduli space is written as
\eqn\vacmodu{
\lf\{\eqalign{
&xy=(u-v)\prod_{a=2}^k(u+c_a)~, \cr
&zw=v\prod_{i=2}^n(v+d_i)~.
}\ri.
}
This moduli space becomes singular when we set some of
the FI-parameters to zero.
The most singular case occurs when all FI parameters are zero.
In this case, the moduli space becomes
\eqn\singvacmodu{
\lf\{\eqalign{
&xy=(u-v)u^{k-1}~, \cr
&zw=v^n~.
}\ri.
}

To see the nature of the singularity of \singvacmodu,
let us recall the case of 4-dimensional $A_{k-1}$ singularity
described by the equation
\eqn\ctwozkeq{
xy=u^k.
}
This equation can be parametrized by the two complex numbers 
$z_1,z_2\in{\Bbb C}$
\eqn\xyuinAB{
x=z_1^k,\quad y=z_2^k,\quad u=z_1z_2.
}
This parametrization of the variety \ctwozkeq\ by $(z_1,z_2)\in{\Bbb C}^2$
is $k$ to 1, hence we have to mod out by the ${\Bbb Z}_k$ identification
\eqn\idenkAB{
(z_1,z_2)\sim (e^{2\pi i\o k}z_1,e^{-{2\pi i\o k}}z_2).
}
Therefore, \ctwozkeq\ has ${\Bbb C}^2/{\Bbb Z}_k$ singularity at the origin.

Now we go back to the analysis of the singularity of \singvacmodu.
Let us first consider the case $n=1$.
Strictly speaking, the $n=1$ case does not 
follow from the quiver gauge theory, since we need two distinguished
nodes in order to connect two $\h{A}$ Dynkin diagrams, which implies
$k,n\geq2$. However, we can formally set $n=1$ in the equation
\singvacmodu\ without asking where it comes from.
When $n=1$, the moduli space \singvacmodu\ becomes
\eqn\noneeq{
xy=(u-zw)u^{k-1}.
}
When $zw\not=0$, there is a ${\Bbb C}^2/{\Bbb Z}_{k-1}$
singularity at $x=y=u=0$. When $z$ or $w$ vanishes,
the singularity at $x=y=u=0$ is enhanced to
${\Bbb C}^2/{\Bbb Z}_{k}$.
Let us consider the singularity at the origin $x=y=z=w=0$. 
In analogy with the 
$A_{k-1}$ ALE space reviewed in the previous paragraph, we
parametrize \noneeq\ as
\eqn\noneABpara{
x=z_1^k,\quad y=z_2^k,\quad z=z_1z_4,\quad w=z_2z_3,\quad
u=z_1z_2t.
}
Then the equation \noneeq\ becomes
\eqn\tABeq{
1=(t-z_3z_4)t^{k-1}.
} 
Since this space is regular, \tABeq\ does not introduce any constraint
on the variables $z_3$ and $z_4$. Therefore, the space \noneeq\
is parametrized by $(z_1,z_2,z_3,z_4)\in{\Bbb C}^4$ with the identification
\eqn\ZkonABAB{
(z_1,z_2,z_3,z_4)
\sim (e^{{2\pi i\o k}}z_1,e^{-{2\pi i\o k}}z_2,e^{{2\pi i\o k}}z_3,
e^{-{2\pi i\o k}}z_4).
}
Namely, the space \noneeq\ has the orbifold singularity 
${\Bbb C}^4/{\Bbb Z}_k$ at the origin.

Similarly, we can  analyze the singularity of \singvacmodu\
for the case $n\geq2$ by rewriting $(x,y,z,w,u,v)$ as
\eqn\ZnZkinAB{
x=z_1^k,\quad y=z_2^k,\quad
z=(z_1z_4)^n,\quad w=(z_2z_3)^n,\quad
u=z_1z_2t,\quad v=z_1z_2z_3z_4.
}
Again, the equation for the moduli space \singvacmodu\ reduces to the
regular equation \tABeq.
Therefore, the moduli space \singvacmodu\ is parametrized by
$(z_1,z_2,z_3,z_4)\in{\Bbb C}^4$ with the identification
\eqn\ZnZkident{\eqalign{
{\Bbb Z}_k:\quad & (z_1,z_2,z_3,z_4)\sim (e^{{2\pi i\o k}}z_1,e^{-{2\pi i\o k}}z_2,e^{{2\pi i\o k}}z_3,e^{-{2\pi i\o k}}z_4),\cr
{\Bbb Z}_n:\quad & (z_1,z_2,z_3,z_4)\sim (z_1,z_2,e^{{2\pi i\o n}}z_3,
e^{-{2\pi i\o n}}z_4).
}}
Namely, the moduli space \singvacmodu\
has the orbifold singularity 
${\Bbb C}^4/({\Bbb Z}_k\times{\Bbb Z}_n)$ at the origin. 
The moduli space \vacmodu\ with generic FI parameters $c_a,d_i\not=0$
can be thought of 
as a hyperK\"{a}hler resolution of the orbifold
${\Bbb C}^4/({\Bbb Z}_k\times{\Bbb Z}_n)$.

\vskip5mm
\noindent
\centerline{\bf Acknowledgment}
This work is supported in part by
MEXT Grant-in-Aid for Scientific Research \#19740135.

\listrefs
\bye